\documentclass[a4paper,aps,prl,showpacs,twocolumn,superscriptaddress]{revtex4} 

\usepackage[utf8]{inputenc}  
\usepackage[british]{babel}  
\usepackage{graphicx} % Package to insert exteral figures
\usepackage{pgfplots}  %Useful to plot graphs with latex
\usepackage{multirow}

\usepackage{graphicx}  % needed for figures
\usepackage{dcolumn}   % needed for some tables
\usepackage{bm}        % for math
\usepackage{amssymb}   % for math
\usepackage{amsmath}
\usepackage{mdframed}
\usepackage{theorem}
\DeclareMathOperator{\tr}{tr}

\newcommand{\ket}[1]{\ensuremath{|#1\rangle}}

\newcommand{\bra}[1]{\ensuremath{\langle#1|}}

\newcommand{\braket}[2]{\ensuremath{\langle #1|#2\rangle}}
\newcommand{\ketbra}[1]{\ensuremath{| #1 \rangle \!\langle #1 |}}
\newcommand{\abs}[1]{|#1|}

\newcommand{\WW}{\ensuremath{\mathcal{W}}}
\newcommand{\DD}{\ensuremath{\mathcal{D}}}

\newcommand{\ba}{\begin{eqnarray}}
\newcommand{\ea}{\end{eqnarray}}
\newcommand{\ban}{\begin{eqnarray*}}
\newcommand{\ean}{\end{eqnarray*}}
\newcommand{\be}{\begin{equation}}
\newcommand{\ee}{\end{equation}}

\begin{document}
\nonfrenchspacing
\title{Characterizing Genuine Multilevel Entanglement}

\author{Tristan Kraft}
\thanks{These two authors contributed equally.}
\affiliation{Naturwissenschaftlich-Technische 
Fakultät, Universität Siegen, Walter-Flex-Straße 3, 57068 Siegen, Germany}
\author{Christina Ritz}
\thanks{These two authors contributed equally.}
\affiliation{Naturwissenschaftlich-Technische 
Fakultät, Universität Siegen, Walter-Flex-Straße 3, 57068 Siegen, Germany}
\author{Nicolas Brunner}
\affiliation{D\'epartement de Physique Th\'eorique, Universit\'e de Gen\`eve, 1211 Gen\`eve, Switzerland}
\author{Marcus Huber}
\affiliation{Institute for Quantum Optics and Quantum Information,
Austrian Academy of Sciences, A-1090 Vienna, Austria}
\author{Otfried Gühne}
\affiliation{Naturwissenschaftlich-Technische 
Fakultät, Universität Siegen, Walter-Flex-Straße 3, 57068 Siegen, Germany}

\date{\today}  

%, 37.10.Ty}
%03.65.Ta: Foundations of quantum mechanics; measurement theory
%03.65.Ud: Entanglement and quantum nonlocality
%(e.g. EPR paradox, Bell's inequalities, GHZ states, etc.)
%42.50.Xa: Optical tests of quantum theory
%37.10.Ty: Ion trapping

\begin{abstract}
Entanglement of high-dimensional quantum systems has become increasingly
important for quantum communication and experimental tests of nonlocality.
However, many effects of high-dimensional entanglement can be simulated 
by using multiple copies of low-dimensional systems. We present a 
general theory to characterize those high-dimensional quantum states 
for which the correlations cannot simply be simulated by low-dimensional 
systems. Our approach leads to general criteria for detecting
multilevel entanglement in multiparticle quantum states, which 
can be used to verify these phenomena experimentally. 
\end{abstract}
\maketitle

{\it Introduction.---}
Entangled quantum systems are now routinely prepared and manipulated 
in labs all around the world, using all sorts of physical platforms. 
In particular, there has been tremendous progress for creating high-dimensional 
entangled systems, which can in principle contain a very large amount of 
entanglement \cite{krenn2014, howell2016, martin2017}. This makes such 
systems interesting from the perspective of information processing, as 
they can enhance certain protocols in particular in quantum communications 
\cite{bechmann2000, cerf2002}. At first sight, the tools of entanglement 
theory can readily be applied to experiments generating high-dimensional 
entangled states. After a closer look, however, one realizes that 
this is not the case in general. Let us illustrate our argument via 
a simple example.

\begin{figure}[t]
    \begin{center}
    \includegraphics[width=0.75\columnwidth]{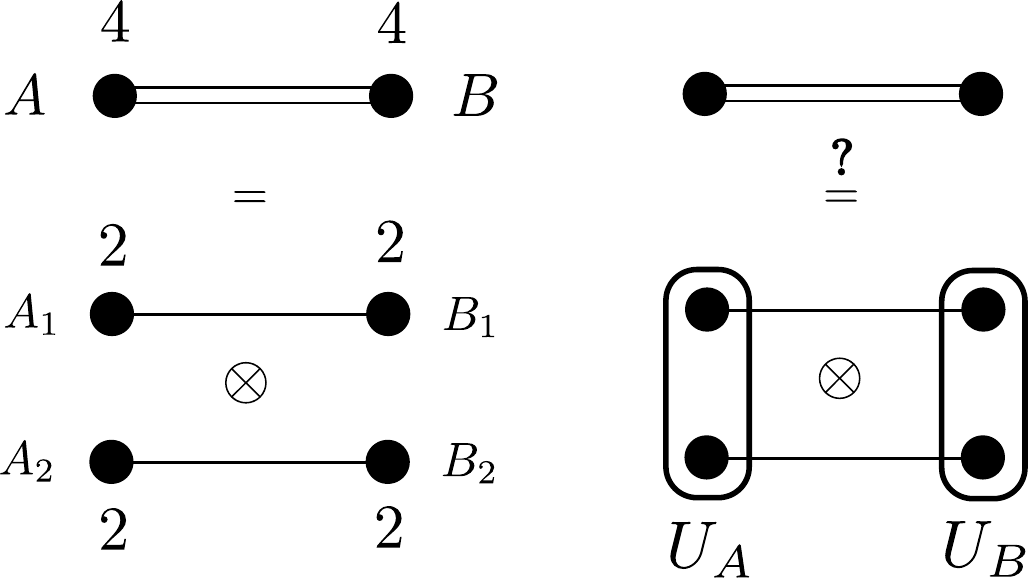}
    \end{center}
    \caption{Left: The four-dimensional maximally entangled state $\ket{\psi_4}$ 
    shared by the parties $A$ and $B$ directly decomposes in two entangled pairs 
    of qubits shared by $A_1 B_1$ and $A_2 B_2$. 
    Right: More generally, we ask whether a high-dimensional entangled 
    state can be decomposed into pairs of entangled systems of smaller dimension, 
    up to some local unitary operations. We show that this is not
    always possible and characterize those states carrying genuine 
    multilevel entanglement.}
    \label{fig:figure1}
\end{figure}

Imagine an experimentalist who wants to demonstrate his ability to 
entangle two high-dimensional quantum systems. He decides to prepare 
the optimal resource state, the maximally entangled state, in increasingly
large dimensions. First, he successfully entangles two qubits in the 
state 
$\ket{\psi_2}=(\ket{00}+\ket{11})/\sqrt{2}$ and two qutrits in the 
state $\ket{\psi_3}=(\ket{00}+\ket{11}+\ket{22})/\sqrt{3}$. 
While preparing the two ququart maximally entangled state 
$\ket{\psi_4}=(\ket{00}+\ket{11}+\ket{22}+\ket{33})/2$ he realizes 
that he could also prepare the two-qubit Bell state $\ket{\psi_2}$ 
twice, see Fig.~\ref{fig:figure1}(a). Clearly, the two copies are equivalent 
to the maximally entangled state of two ququarts when identifying 
$\ket{00}_{A_1 A_2} \mapsto \ket{0}_A$, 
$\ket{01}_{A_1 A_2} \mapsto \ket{1}_A$, 
$\ket{10}_{A_1 A_2} \mapsto \ket{2}_A$,
and
$\ket{11}_{A_1 A_2} \mapsto \ket{3}_A$.
Furthermore, using the source $n$ times, the experimentalist prepares 
the state $\ket{\psi_2}^{\otimes n}$, which is equivalent to a maximally 
entangled state in dimension $2^n \times 2^n$. The experimentalist is 
thus enthusiastic, as he now has access to essentially any entangled 
state with an entanglement cost of at most $n$ ebits. In particular 
this should allow him to implement enhanced quantum information protocols 
based on high-dimensional entangled states, which are proven to boost 
the performance of certain protocols. 

Clearly, the view of the experimentalist is too simplistic and key aspects have been 
put under the carpet. In order to use the full potential of the state, 
and thus really claim to have access to high-dimensional entanglement, the experimentalist 
should be able to perform arbitrary local measurements, including joint measurements between the two subspaces (e.g., photons), which can be non-trivial 
to implement in certain experimental setups. Ideally, the experimentalist 
should be able to implement arbitrary local transformations on the local 
four-dimensional space.

If one focuses on the generated state, however, the known methods of 
entanglement verification support the naive view of the 
experimentalist. For instance, there are tools to certify the Schmidt rank
of the state \cite{sw1, sw2}, but these do not distinguish between many copies of
a Bell state and a genuine high-dimensional state. Bell inequalities 
have been proposed as dimension witnesses for quantum systems \cite{dimwit}, but 
recently it has turned out that these do not recognize the key feature, 
as independent measurements on two Bell pairs can mimic the statistics
of a high-dimensional system \cite{yuthesis, cong}. So they just
characterize the Schmidt rank in a device-independent manner.

In this work, we characterize the high-dimensional quantum states 
which give rise to correlations that can not be simulated many 
copies of small-dimensional systems. This leads to the notion of
genuine multilevel entanglement and we show how this can be created 
and certified. Then we extend this idea to the multiparticle case. 
Our results imply that many of the prominent entangled states in 
high dimensions can directly be simulated with small-dimensional 
systems.

%%%%%%%%%%%%%%%%%%%%%%%%%%%%%%%%%%%%%%%%%%%%%%%%%%%%%%%%%%%%%%%%%%

{\it The scenario.---}
First we consider four-level systems, also called ququarts. A general two-ququart 
entangled state can be written in the Schmidt decomposition as
\begin{equation}
\ket{\psi}
=
s_0\ket{00}_{AB}+s_1\ket{11}_{AB}+s_2\ket{22}_{AB}+s_3\ket{33}_{AB},
\label{Eq:State}
\end{equation}
where we assume here and in the following the Schmidt coefficients to be ordered, 
i.e., $s_0\geq s_1\geq s_2\geq s_3\geq 0$ and $\sum_i s_i^2=1$. One can replace 
each ququart with two qubits, so the total state may also be considered as
a four-qubit state. The question we ask is whether it is possible to reproduce 
any correlations in the two-ququart state by preparing two entangled pairs of 
qubits only (see Fig.~\ref{fig:figure1}).

A first approach is to replace on Alice's side 
$\ket{0}\mapsto \ket{00}$, 
$\ket{1}\mapsto \ket{01}$,  $\ket{2}\mapsto \ket{10}$, 
and $\ket{3}\mapsto \ket{11}$ and similarly for  Bob. 
Note that this is, so far, 
not guaranteed to be the optimal assignment of basis states on two 
qubits to the basis states $\{\ket{0}, \ket{1}, \ket{2},\ket{3}\}.$
This replacement leaves us  with the 
four-qubit state
\begin{eqnarray}
\ket{\psi}&=& s_0\ket{00}_{A_1 B_1}\ket{00}_{A_2 B_2}+s_1\ket{00}_{A_1 B_1}\ket{11}_{A_2B_2} 
\notag \\
&+&s_2\ket{11}_{A_1 B_1}\ket{00}_{A_2 B_2}+s_3\ket{11}_{A_1 B_1}\ket{11}_{A_2 B_2}.
\end{eqnarray}
Now we ask under which conditions this state 
can be decomposed as
\begin{eqnarray}
\ket{\varphi} &=&
(\alpha_0\ket{00}_{A_1 B_1}+\alpha_1\ket{11}_{A_1 B_1})
\notag \\
&\otimes & (\beta_0\ket{00}_{A_2 B_2}+\beta_1\ket{11}_{A_2 B_2}).
\end{eqnarray}
For the Schmidt coefficients it must hold that 
$s_0=\alpha_0\beta_0$, $s_1=\alpha_0\beta_1$, $s_2=\alpha_1\beta_0$ and 
$s_3=\alpha_1\beta_1$. If $\ket{\psi}$ can be written in this form, we call 
$\ket{\psi}$ decomposable and otherwise genuinely four-level entangled.
An interesting example is the maximally entangled state of two ququarts, 
$\ket{\psi_4}= 
(\ket{00}_{AB} +\ket{11}_{AB}+\ket{22}_{AB}+\ket{33}_{AB})/2$. 
Here $s_i={1/2}$ and for $\alpha_0=\alpha_1=\beta_0=\beta_1={1/{\sqrt{2}}}$ 
we have $\ket{\psi}=\ket{\varphi}$. Thus the maximally entangled state is decomposable, its correlations are reproducible by two pairs of entangled 
qubits, and the state is not sufficient to certify genuine four-level 
entanglement.

In order to decide decomposability for a general $\ket{\psi}$ we compute the maximal 
overlap between $\ket{\psi}$ and all decomposable states $\ket{\varphi}$:
\begin{align}
\label{maxSV}
\max_{\ket{\phi}}
&
\abs{\braket{\psi}{\varphi}}=\max_{\alpha_i,\beta_i}\,\{s_0\alpha_0\beta_0+s_1\alpha_0\beta_1+s_2\alpha_1\beta_0+s_3\alpha_1
\beta_1\}
\notag \\
&
=\max_{\alpha,\beta} \bra{\beta}S\ket{\alpha}
=\max\,\text{singval}(S),
\end{align}
where $\ket{\alpha}=(\alpha_0,\alpha_1)^T$,
$\ket{\beta}=(\beta_0,\beta_1)^T$,
and
\begin{equation}
S= 
\begin{bmatrix}
s_0&s_1\\s_2&s_3
\end{bmatrix},
\end{equation}
and $\text{singval}(S)$ denotes the singular values.

Note that up to now, we have not determined the optimal choice for the basis assignment,
that is, we used the simple assignment $\ket{0}\mapsto \ket{00}$ etc.~introduced above. The optimal assignment can be determined by optimizing 
over local unitaries on the ququarts. In Appendix A \cite{appremark}
we show that the maximal singular value 
is obtained if the states $\ket{\psi}$ and $\ket{\varphi}$ have the same Schmidt basis and 
the remaining freedom encompasses permutations in the assignment of basis elements. As it 
turns out, the basis choice from the beginning is optimal and we have:

\noindent
{\bf Observation 1.} 
{\it 
The two-ququart state $\ket{\psi}$ is decomposable if and only 
if $\max\,\textit{\normalfont{singval}}(S)=1$. This is equivalent 
to $\det(S)=0$. The proof is given in Appendix A \cite{appremark}.
}

The extension of decomposability to mixed states is straightforward. We define a
mixed state to be  decomposable, if it can be written as $\varrho=\sum_i p_i \ketbra{\psi_i}$
where the $\ket{\psi_i}$ are decomposable, and genuine four-level entangled otherwise.
The set of decomposable states $\DD$ is convex by definition. This allows to construct
witnesses for four-level entanglement. Recall that an operator $\WW$ is called an 
entanglement witness, iff $\tr(\sigma \WW)\geq 0$ for all separable states 
$\sigma$ and $\tr(\varrho \WW)<0$ for at least one  entangled state $\varrho$ 
\cite{Terhal_Witness}. A special type of witnesses are the projector-based 
witnesses which are of the form $\WW=\alpha \openone -\ketbra{\xi}$, where $\alpha$ is the 
maximal squared overlap between $\ket{\xi}$ and the decomposable states \cite{gtreview}. In order to detect 
as many states as possible, we chose $\ket{\xi}$ to be the state with the largest 
distance to $\DD$, meaning that $\alpha$ is as small as possible. The state $\ket{\xi}$ 
can be determined by minimizing the maximal singular value of $S$ in Eq.~\eqref{maxSV}
which is, according to Observation 1, a function of squared determinant. Thus we distinguish 
between positive and negative values of the determinant, giving two interesting states 
$\ket{\xi_i}$, see Appendix B \cite{appremark} for details:

\noindent
{\bf Observation 2.}
{\it 
The following two states locally maximize the distance 
to the decomposable states:  
For $\det(S) < 0$ the Schmidt-rank three state
\begin{equation}
\ket{\xi_1}={1 \over {\sqrt{3}}}(\ket{00}+\ket{11}+\ket{22})
\end{equation}
has the largest distance  with $\alpha= [(3+\sqrt{5})/6]^{1 \over 2}\simeq 0.934$ 
to the set of decomposable states. For $\det(S) > 0$ the Schmidt-rank four state 
\begin{equation}
\ket{\xi_2}=\sqrt{{3 \over 4}}\ket{00}+{1 \over {2\sqrt{3}}} (\ket{11}+\ket{22}+\ket{33})
\end{equation}
maximizes the distance with a value of $\alpha= [(3+2\sqrt{2})/6]^{1 \over 2}\simeq 0.986$ 
to the set of decomposable states.
}

%%%%%%%%%%%%%%%%%%%%%%%%%%%%%%%%%%%%%%%%%%%%%%%%%%%%%%%%%%%%%%%%%%%%%%%%%

{\it General theory for bipartite systems.---}
Let us start by considering only decompositions into two lower-dimensional 
states. In this case the results from the previous section still hold, only 
the matrix $S$ increases according to the dimensions of the subsystems. 
This leaves us with the problem that the maximal singular value depends 
on the encoding, which defines the arrangement of Schmidt coefficients 
in the matrix $S$. 

As an example we consider the embedding of the rank-four state from 
Eq.~\eqref{Eq:State} in a $6\times 6$ dimensional system, that is, 
each party has a qubit and a qutrit. 
Using the encoding $ \ket{0} \mapsto \ket{00}, \ket{1} \mapsto \ket{01}, 
\ket{2} \mapsto {02} , \ket{3} \mapsto \ket{10}, \ket{4} \mapsto \ket{11} , \ket{5} \mapsto \ket{12}$ 
we obtain the matrix $S_1$ whereas using $ \ket{0} \mapsto \ket{00}, \ket{1} \mapsto \ket{01}, 
\ket{2} \mapsto {10}, \ket{3} \mapsto \ket{11}, \ket{4} \mapsto \ket{02} , \ket{5} \mapsto \ket{12}$ 
we obtain a different matrix $S_2$. The matrices are given by
\begin{equation}
S_1=
\begin{bmatrix}
s_0 & s_1 & s_2 \\ s_3 & s_4 & s_5
\end{bmatrix},
\qquad 
S_2= 
\begin{bmatrix}
s_0 & s_1 & s_4 \\ s_2 & s_3 & s_5
\end{bmatrix}
\label{differentsmatrices}
\end{equation}
and can lead to different singular values. For instance, if we embed 
the two-ququart state $\ket{\psi_4}$ in this configuration, i.e., $s_0=s_1=s_2=s_3=1/2$ and 
$s_4=s_5=0$, we find that $\text{max singval}(S_1)\neq 1$, whereas 
$\text{max singval}(S_2)= 1$.
Consequently, 
when deciding decomposability, it is crucial to optimize over all 
possible permutations of entries in $S$. As the number of permutations 
grows super-exponentially, it is in general hard to compute this for 
increasing dimensions.

Nevertheless, the complexity can be reduced, as we have to consider 
only those permutations which lead to different maximal singular 
values. In Appendix C \cite{appremark}) we discuss this simplification
which leads to the theory of Young tableaux \cite{Georgi}. It turns out 
that for a decomposition into $d=d_1\times d_2$ there are at most
\begin{equation}
\mathcal{N}=\frac{(d_1\times d_2)!}{\prod_{i=1}^{d_1} \prod_{j=1}^{d_2}(i+j-1)}
\end{equation}
different matrices that could lead to different singular values. Examples 
can be found in Appendix B \cite{appremark}.

Furthermore, if one is only interested in decomposability, it suffices 
to check whether there exists an arrangement such that $S$ has rank one. 
The number of possible arrangements reduces to at most
\begin{equation}
\mathcal{N'}=\frac{(d_1 + d_2 -2 )!}{(d_1-1)! \times (d_2-1)!} . %hook length
\end{equation}
It should be noted that an equivalent problem and solution has 
been considered in quantum thermodynamics, where one may ask whether 
the correlations in a bipartite system can drop to zero under global 
unitaries \cite{jevtic}. 

To complete the discussion, 
one may also take into account a decomposition of the system into more than
two lower-dimensional subsystems. In this case, the matrix $S$ becomes
a tensor and thus deriving an analytical expression, equivalent to the 
singular value decomposition, is difficult. However, one can construct
an iterative algorithm to calculate the maximal overlap between the 
original state and a given set of decomposable states as follows: The 
total maximization can be split into a maximization over states and 
local unitaries. If all but one of these objects are fixed, the remaining
one can be carried out analytically. This leads to a fast iteration, see 
Appendix E \cite{appremark} for a detailed discussion.

%%%%%%%%%%%%%%%%%%%%%%%%%%%%%%%%%%%%%%%%%%%%%%%%%%%%%%%%%%%%%%%%%%%%
{\it Multiparticle systems.---}
We call an $N$-partite pure state $\ket{\psi}$ in $(\mathbb{C}^{D})^{\otimes N}$ 
fully decomposable iff there exist $N$-partite states $\ket{\varphi}$, $\ket{\varphi'}$ of dimension $d,d'$ such that:
\begin{equation}
\label{eq:dec}
\ket{\psi} = U_1\otimes\cdots\otimes U_N\ket{\varphi} \otimes \ket{\varphi'},
\end{equation}
for some $d\times d'=D$. Here, the $U_i$ denote the unitaries each party 
applies to their local subsystems. This definition is in analogy 
to full separability in entanglement theory \cite{gtreview}. A state that is not 
fully decomposable is multipartite multilevel entangled 
(MME). 

If a state is non-decomposable according to Eq.~\eqref{eq:dec}, there 
might exist partitions under which such states are decomposable. For instance, 
a state may be decomposable, if the unitary on the first two particles is 
allowed to be nonlocal, i.e., we may set 
$U_1 \otimes U_2 \mapsto U_{12}^{\rm nl}.$
More generally, there may be a bipartition of 
the $N$ particles for which the state is decomposable.

\noindent
{\bf Observation 3.}
{ \it
Consider an $N$-particle state $\ket{\psi}$.
If there exists a bipartition $M|M'$ of the $N$ particles 
for which the state is decomposable, the state is called bidecomposable. 
Otherwise the state is genuinely multipartite multilevel 
entangled (GMME). Verifying GMME for pure states can be done
by applying the methods for bipartite systems to all bipartitions.
}

To show that a pure multiparticle state is not fully decomposable is, 
however, not straightforward,  as there is in general no Schmidt 
decomposition for systems consisting of more than two parties \cite{peresold}.
Nevertheless, the iterative algorithm mentioned above can 
again be utilized. Note that within the optimally decomposed state, the 
largest block that cannot be decomposed any further identifies the 
minimal number of parties and dimensions needed to reproduce the 
correlations in the original state. Also, the definitions above can
be readily generalized to mixed states by considering convex combinations.
In the following sections, we discuss examples which are relevant
for current experiments. 

%%%%%%%%%%%%%%%%%%%%%%%%%%%%%%%%%%%%%%%%%%%%%%%%%%%%%%%%%%%%%%%%%%%%%%

{\it Example 1: Generalized GHZ states.---}
Motivated by our result from the bipartite case that the 
maximally entangled state is decomposable, we start with 
studying Greenberger-Horne-Zeilinger (GHZ) states,
$\ket{GHZ^{(D)}}={1 \over{\sqrt{D}}} (\ket{0\cdots 0}+\ket{1\cdots 1}+\cdots +\ket{(D-1) \cdots (D-1)})$ for $N$ particles with local dimension 
$D$. 

First, we observe that the GHZ state is fully decomposable. In fact,
it is decomposable with respect to the finest factorization of the 
local dimension $D$, given by the prime decomposition $D=\prod_{j=1}^k d_{j}$ 
of $D$, as we can write:
\begin{equation} 
\label{GHZ}
\ket{GHZ^{(D)}} \stackrel{enc.}{=}
\bigotimes \nolimits_{j=1}^{k} \ket{GHZ^{(d_j)}},
\end{equation}
where $\ket{\varphi_j}$ represents the $N$-partite state of the 
subsystem with dimension $d_j$.

The proof  is straightforward. We just replace 
each level $\ket{i}$ (with $ i \in [0,D-1]$) of the original state with 
its respective encoding into the lower levels $\ket{i_1,\dots,i_k}$ where 
each $i_j$ has dimension $d_j$ and as such values $\in [0,d_j-1]$  for all 
$j$. The ordering of the encoding is chosen such that the value within the 
respective number system is increasing, that is it corresponds to a binary 
encoding for qubits ($d_j=2$), ternary for qutrits ($d_j=3$), and similarly 
for higher dimensions. This leads to 
$\ket{0} \mapsto \ket{0 \dots 0},..., 
\ket{D-1} \mapsto \ket{\bigotimes_j (d_j-1),\dots, \bigotimes_j (d_j-1)}$. Following this encoding process, a reordering, that is $(A_1\dots A_n,B_1\dots B_n,\dots) \mapsto (A_1B_1\dots,\dots,A_nB_n\dots)$, directly reveals the tensor structure of the encoded state with respect to every factor $d_j$.
In Appendix D \cite{appremark} we give the calculation for a six-dimensional 
GHZ state.  Furthermore we show there that the absolutely maximally entangled 
state of six qubits represents a decomposable three-ququart state in
the GHZ class.

So all the correlations of a GHZ state in high dimensions, although having a high 
Schmidt-rank for the bipartitions, can be simulated by low-dimensional systems. This 
is distinct from other approaches, such as the Schmidt number vectors from Ref.~\cite{hubervicente} 
or the criterion in Ref.~\cite{old_chinese_paper}, where the GHZ state was used to detect 
higher-order entanglement. For completeness, a proof of the LU-equivalence between GHZ- and 
the star-type graph states from Ref.~\cite{old_chinese_paper} is given in Appendix D \cite{appremark}.

%%%%%%%%%%%%%%%%%%%%%%%%%%%%%%%%%%%%%%%%%%%%%%%%%%%%%%%%%%%%%%%%%
\begin{figure}[t]
 \centering
 \includegraphics[width=0.8\columnwidth]{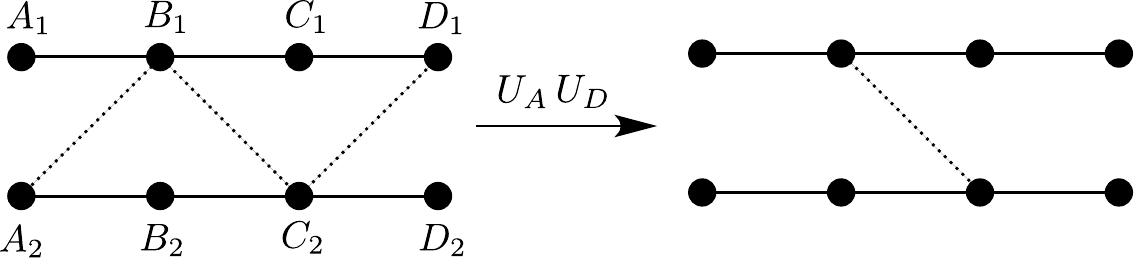}
 \caption{Examples of weighted graph states. Left: 
 The four-ququart chain-graph state from Eq.~(\ref{QQchain})
 can be encoded into a weighted graph state of eight qubits, see 
 Eq.~(\ref{QBwe}). Right: After application of the unitaries $U_{A_1 A_2}$ 
 and $U_{D_2 D_1}$ the state exhibits decomposability with respect to 
 the bipartitions $A|BCD$, $D|ABC$ and $AD|BC$ [see Eq.~(\ref{bidecgraph})]
 and thus the original ququart state is bidecomposable and not GMME.}
 \label{fig:mme}
\end{figure}
%%%%%%%%%%%%%%%%%%%%%%%%%%%%%%%%%%%%%%%%%%%%%%%%%%%%%%%%%%%%%%%%%

{\it Example 2. Graph states.---}
A $D$-dimensional weighted graph state  can be written 
as \cite{q.s.with_qudit_graphs, weighted_graphs_hein}
\begin{eqnarray}
\ket{G}= \prod\nolimits_{\{ij\} \in E} Z_{\{ij\}}^{\alpha} \ket{+^D}^{\otimes~V},
\label{eq:graphstatedef}
\end{eqnarray}
where $V$ denotes the set of vertices, $E$ the set of edges connecting 
two vertices $i$ and $j$ and $\ket{+^D}$ is given by
$\ket{+^D} \propto \ket{0}+\ket{1}+...+\ket{D-1}$. Entanglement is created 
by the controlled Z-gates
$Z_{\{ij\}}^{\alpha}=\sum_{g=0}^{d-1} (\ket{g} \bra{g})_i\otimes Z_{j} ^{g \alpha}$, 
where $Z_q=\sum_{q=0}^{D-1} \omega^{q} \ket{q} \bra{q}$ 
(with $\omega=e^{{2\pi i}/{D}}$) defines the single-qudit Z-gate. 
For $\alpha =1$ the structure reduces to non-weighted graph states, 
for $\alpha={1 \over 2}$ the weighted edges can be graphically represented by dashed lines.

As an example for a state which is MME but not GMME, consider 
the chain graph state of four ququarts: 
\begin{equation} 
\label{QQchain}
\ket{G^{(4)}}= Z_{AB}Z_{BC}Z_{CD}\ket{+^4}^{\otimes 4}.
\end{equation}
Encoding to eight qubits gives us the state (see Fig.~\ref{fig:mme}, 
detailed calculations can be found in Appendix D \cite{appremark}): 
\begin{align}
\ket{G^{(2)}} 
&=Z_{A_1 B_1} Z_{B_1 C_1} Z_{C_1D_1} 
Z_{A_2 B_2} Z_{B_2 C_2} Z_{C_2 D_2}
\nonumber
\\
 & \times Z_{A_2 B_1}^{1\over 2} Z_{B_1C_2}^{1\over 2} Z_{C_2D_1}^{1\over 2}\ket{+^2}^{\otimes 8}.
 \label{QBwe-null}
\end{align}
Now we apply two-qubit unitaries of the form
$
U_{ij}=
\ket{+}\bra{+}_i \otimes \openone_j + \ket{-}\bra{-}_i \otimes Z^{3/2}_j
$
with $\ket{\pm}=(\ket{0}\pm \ket{1})/\sqrt{2}$
on the two qubits of system A and D respectively and end up with
\begin{align}
\label{QBwe}
&
U_{A_1A_2} U_{D_2D_1} \ket{G^{(2)}}=Z_{B_1C_2}^{1\over 2}\ket{G^{(2)}_{\DD}},
\mbox{ where }
\\
&\ket{G^{(2)}_{\DD}}= 
Z_{A_1 B_1} Z_{B_1 C_1} Z_{C_1D_1} 
Z_{A_2 B_2} Z_{B_2 C_2} Z_{C_2 D_2}
\ket{+^{2}}^{\otimes 8}
\label{bidecgraph}
\end{align}
is a fully decomposable state with no diagonal edges.

Thus for the bipartitions $A|BCD$ or $D|ABC$ the state is decomposable 
and thereby not GMME. In fact, we find decomposability 
with respect to every possible bipartition (see Appendix D \cite{appremark}).
For claiming multilevel entanglement, we still have to exclude full decomposability,
which is, as mentioned, a difficult task. We applied a numerical 
algorithm (Appendix E \cite{appremark}) which strongly indicates 
non-decomposability with an maximal overlap of $0.8536$ with the set of 
fully decomposable states.

%%%%%%%%%%%%%%%%%%%%%%%%%%%%%%%%%%%%%%%%%%%%%%%%%%%%%%%%%%%%%%%

{\it Example 3. A genuine multilevel entangled state.---} 
As a final example, consider the three ququart state
\begin{equation}
\ket{\psi^{(4)}}=
\sum_{j=0}^{3} \ket{u_j}\ket{j}\ket{u_j} -2\ket{3}\ket{3}\ket{3},
\label{psi4}
\end{equation}
where  $\ket{u_0}=\ket{0}+\ket{1}+\ket{2}+\ket{3}$, 
$\ket{u_1}=\ket{0}-\ket{1}+\ket{2}-\ket{3}$, 
$\ket{u_2}=\ket{0}+\ket{1}-\ket{2}-\ket{3}$, 
$\ket{u_3}=\ket{0}-\ket{1}-\ket{2}+\ket{3}$.
This state corresponds to the six-qubit state $\ket{\psi^{(2)}}= Z_{123456} Z_{13} Z_{35} Z_{24} Z_{46} \ket{+^{(2)}}$,
a graph state with an additional hyperedge connecting all vertices \cite{hypergraphs}.
For this state we found for all bipartitions the Schmidt coefficients to 
be $s_0= 0.551$, $s_1=s_2=0.5$, $s_3=0.443$ which leads to a non-zero 
determinant of $\det(S)= -0.0059$.
Hence, ${\rm rank}(S)\neq 1$ for all bipartitions and the state is 
non-decomposable for any bipartition. So the state is GMME, to be exact, genuine 3-partite 4-level entangled.
 
%%%%%%%%%%%%%%%%%%%%%%%%%%%%%%%%%%%%%%%%%%%%%%%%%%%%%%%%%%%%%%%%% 

%%%%%%%%%%%%%%%%%%%%%%%%%%%%%%%%%%%%%%%%%%%%%%%%%%%%%%%%%%%%%%%%% 

{\it Conclusion.---}
We have introduced the notion of genuine multilevel entanglement.
This formalizes the notion of high-dimensional entanglement that 
cannot be simulated directly with low-dimensional systems.  We have 
provided methods to characterize those states for the bipartite and 
multipartite case, including the construction of witnesses for an 
experimental test. The results can be interpreted as a cautionary tale 
with regards to naively employing standard entanglement characterization 
tools. Whereas under general local operations 
and classical communication, multiple copies of small dimensional systems
are universal, this is not the case anymore in restricted scenarios, even 
having access to all possible local unitaries. This suggests that
high-dimensional quantum systems do present a fundamentally different resource 
under realistic conditions.

For future research there are different topics to address. First, one 
may consider network scenarios, where a high-dimensional quantum state is 
distributed between several parties, and the correlations should be explained
by low-dimensional states shared between subsets of the parties. Second, it 
would be desirable to develop a resource theory of high-dimensional entanglement, 
where not only the state preparation, but also the local operations (like 
filters) of the parties are considered. This may finally lead to a full 
understanding of quantum information processing with high-dimensional systems.

We thank Yu Cai,  Wan Cong, and Valerio Scarani %, and Petronilla Benasquez 
for discussions. This work has been supported by the ERC (Consolidator 
Grant 683107/TempoQ), the DFG, the Swiss National Science Foundation 
(Starting grant DIAQ and QSIT, and the Austrian Science Fund (FWF) through 
the START project Y879-N27 and the international project I3053-N27.

%%%%%%%%%%%%%%%%%%%%%%%%%%%%%%%%%%%%%%%%%%%%%%%%%%%%%%%%%%%%%%%%%

\section*{Appendix}

\subsection{A: Proof of Observation 1}
Here we prove Observation 1, which states that a two ququart 
state is decomposable iff $\text{max\,sinval}(S)=1$, where
\begin{equation}
S=
\begin{bmatrix}
 s_0&s_1\\s_2&s_3
\end{bmatrix}.
\end{equation}
First, let us consider two bipartite ququart states $\ket{\psi}$ and $\ket{\varphi}$. We prove that the maximal overlap between $\ket{\psi}$ 
and $\ket{\varphi}$, where each party is allowed to perform local unitary operations, is given by:
\begin{equation}
F_{max}=\max_{U_A,U_B}\abs{\bra{\psi}U_A\otimes U_B\ket{\varphi}}=\sum_{i=0}^{D-1}
\eta_i\sigma_i
\end{equation}
where $\eta_0\geq\dots\geq\eta_{3}\geq 0$ are the Schmidt coefficients of 
the state $\ket{\psi}$ and $\sigma_0\geq\dots\geq\sigma_{3}\geq 0$ are the 
Schmidt coefficients of the state $\ket{\varphi}$. This was already shown 
in Ref.~\cite{nonlin_entwit}, but we add this here for completeness. We start 
by writing the overlap in terms of coefficient matrices of the states 
$\ket{\psi}$ and $\ket{\varphi}$, that is we write $\ket{\psi}=\sum_{i,j}C_{\psi}^{ij}\ket{ij}$ as $C_{\psi}=\sum_{ij}C^{ij}_{\psi}\ket{i}\bra{j}$, 
and similarly for $\ket{\varphi}$. We have
\begin{eqnarray}
\notag 
F_{max}&=&\max_{U_A,U_B}\abs{\bra{\psi}U_A\otimes U_B\ket{\varphi}} 
\\
\notag 
&=&\max_{U_A,U_B}\abs{\tr(C^{\dagger}_{\psi} U_A C_{\varphi} U_B^T)} 
\\
&=&\sum_{i=0}^{D-1} s_i(C_{\psi})s_i(C_{\varphi}).
\label{Eq:Neumann}
\end{eqnarray}
In the last step of Eq.~\eqref{Eq:Neumann} we used von Neumann’s trace 
inequality:
\begin{equation}
\label{NeumannTr}
|\tr(\Lambda \Gamma)| \leq \sum_i \lambda_i \gamma_i 
\end{equation}
which holds for all complex $n \times n$ matrices $\Lambda$ and $\Gamma$ 
with ordered singular values $\lambda_i \leq \lambda_{i-1}$ and 
$ \gamma_i \leq \gamma_{i-1}$. It was proven in Ref.~\cite{mirsky} 
that equality in  Eq.~\eqref{NeumannTr} can only be reached when 
$\Lambda$ and $\Gamma$ are simultaneously unitarily diagonalizable 
and hence both states need to have the same Schmidt basis. 
Therefore it is optimal to choose the encoding between the 
four-dimensional systems and the qubits in the Schmidt basis.  
Furthermore note that the singular values of the coefficient 
matrices are nothing but the Schmidt coefficients of the state. 
For the $2\times 2$ matrix $S$ the maximal singular value is 
given by
\begin{equation}
\alpha=\sqrt{\frac{1}{2}(1+\sqrt{1-4\det(S)^2})}.
\label{Eq:alpha}
\end{equation}
Hence, we find that $\text{max singval}(S)=1$ iff $\det(S)=0$, which 
finishes the proof of Observation 1. Other encodings lead to the same 
result since changing the encoding, can, for the special case of two qubits,
be described by swapping rows or columns of $S$, which does not change its 
singular values. Note that for higher-dimensional systems (e.g., a qubit 
and a qutrit) the last point is not true, and this is the reason why we have
to consider different matrices $S$ there [see Eq.~(\ref{differentsmatrices})
in the main text].

%%%%%%%%%%%%%%%%%%%%%%%%%%%%%%%%%%%%%%%%%%%%%%%%%%%%%%%%%%%%%%%%

\subsection{B: Witnesses for the bipartite case}
Here we show how to construct a witness operator for four-level entanglement.
We are seeking for the state $\ket{\xi}$ which has the largest distance to the 
set of decomposable states and the smallest coefficient $\alpha$ such that the 
witness $\WW=\alpha\openone-\ketbra{\psi}$ is positive on all decomposable 
states. Note that since $\DD$ is a convex set, it is sufficient to optimize 
over all pure decomposable states. In order to find $\ket{\xi}$ we compute
\begin{align}
  \alpha &= \min_S[\max \,\text{singval}(S)]   \nonumber\\
  &\text{s. t.:}\quad  \det(S) \neq 0,    \nonumber\\
 &\quad \quad s_0^2+s_1^2+s_2^2+s_3^2=1,  \nonumber\\
 &\quad \quad s_0 \geq s_1 \geq s_2 \geq s_3\geq 0.
 \label{sdptest}
\end{align}
First note that the maximal singular value of a $2\times 2$ matrix is 
of the form of Eq.~\eqref{Eq:alpha}. In the following we separately analyse the cases $\det(S) < 0$ and 
$\det(S) >0$. 

For $\det(S)<0$ we have to minimize $\det(S)=s_0s_3-s_1s_2$. Since $s_3$ is by definition the smallest coefficient we choose $s_3=0$. Then we are left with $\max s_1\cdot s_2$. For fixed $s_0$ we have that
\begin{equation}
s_1^2+s_2^2=const.
\label{eq:circle}
\end{equation}
which is the equation of a circle. Therefore the problem is equivalent to maximizing the area of a rectangle with one corner at the origin and the other one on the circle defined by Eq.~\eqref{eq:circle}. The obvious solution is therefore $s_1=s_2$. Since $s_0\geq s_1$ the maximum is obtained at
$s_0=s_1=s_2=\frac{1}{\sqrt{3}}$.

For $\det(S)>0$ we have to maximize $\det(S)=s_0s_3-s_1s_2$. Therefore we 
have for any given $s_0,s_3$ to minimize $f(s_1)=s_1\cdot s_2=s_1\sqrt{C-s_1^2}$ such that $s_0\geq s_1\geq s_2\geq s_3\geq 0$ and $C=1-s_0^2-s_3^2$. The minimum 
of the function $f(s_1)$ is obtained at the boundary for $s_1=s_3$, which implies $s_2=s_3$. Therefore the maximum of the determinant is obtained at $s_1=s_2=s_3=\frac{1}{2\sqrt{3}}$ and $s_0=\sqrt{3/4}$.

We see that for dimension four the state with the largest distance to 
the set of decomposable states is the maximally entangled state of two 
qutrits. We observe that for increasing dimensions the distance between 
the maximally entangled states with lower dimension and the set of decomposable states decreases. Some analytical and numerical values are shown in Table \ref{Tab:Numerical}. This might lead to the conclusion that the multilevel entangled states get closer to the set of decomposable states for larger dimensions. However a proof that the maximally entangled states are the 
ones having the largest distance to the set of decomposable states is still missing.

\begin{table}[t!]
\begin{tabular}{c|c|c}
	Source & rank & overlap\\
    \hline
    \hline
    \multirow{1}{*}{$2\times 2\, (\bf{4})$}& 3 & $\sqrt{\frac{1}{6}(3+\sqrt{5})}\simeq 0.934$\\
    \hline
    \multirow{1}{*}{$2\times 3\, (\bf{6})$}& 5 & $\sqrt{\frac{1}{10}(5+\sqrt{17})}\simeq 0.955$\\
    \hline
    \multirow{2}{*}{$2\times 4\, (\bf{8})$}& 5 & $\sqrt{\frac{1}{10}(5+\sqrt{17})}\simeq 0.955$\\
    & 7 & $\sqrt{\frac{1}{14}(7+\sqrt{37})}\simeq 0.966$\\
    \hline
    \multirow{3}{*}{$3\times 3\, (\bf{9})$}& 5 & $\sqrt{\frac{1}{10}(5+\sqrt{17})}\simeq 0.955$\\
    & 7 & $\sqrt{\frac{1}{14}(7+\sqrt{33})}\simeq 0.954$\\
    & 8 & $\sqrt{\frac{1}{16}(7+\sqrt{48})}\simeq 0.965$\\
    \hline
    \multirow{2}{*}{$2\times 4\, (\bf{10})$}& 7 & $\sqrt{\frac{1}{14}(7+\sqrt{37})}\simeq 0.966$\\
    & 9 & $\sqrt{\frac{1}{18}(9+\sqrt{65})}\simeq 0.973$\\
    \hline
    \multirow{2}{*}{$2\times 6\, (\bf{12})$}& 7 & $\sqrt{\frac{1}{14}(7+\sqrt{37})}\simeq 0.966$\\
    & 9 & $\sqrt{\frac{1}{18}(9+\sqrt{65})}\simeq 0.973$\\
    \hline
	\multirow{1}{*}{$7\times 7\, (\bf{49})$}& 11 & $\sqrt{\frac{1}{22}(11+\sqrt{101})}\simeq 0.9781$\\
    \hline
\end{tabular}
\caption{This table shows the analytical and numerical fidelities of the maximally entangled state $\ket{\psi}=1/\sqrt{D}\sum_{i=0}^{D-1}\ket{ii}$ with all decomposable states for a given dimension of the source.}
\label{Tab:Numerical}
\end{table}

%%%%%%%%%%%%%%%%%%%%%%%%%%%%%%%%%%%%%%%%%%%%%%%%%%%%%%%%%%%%%%%%%%%%%%%%%%%5

\subsection{C: Connection to the theory of Young tableaux}

In this section we want to discuss the relation between the number of 
arrangements of Schmidt coefficients in the matrix $S$ and the number 
of standard Young tableaux.

As mentioned in the main text, the complexity of characterizing the matrices $S$
can be reduced, as we have to consider only those permutations which lead to different 
maximal singular values. First, note that given two probability distributions $\{p_i \}$ 
and $\{q_i\}$ the sum over the products $\sum_i \sqrt{p_i q_i}$ is maximal 
iff both are ordered in the same way. We can further assume in Eq.~(\ref{maxSV}) in 
the main text that $\alpha_0\geq\alpha_1$ and similarly for $\beta_i$, since exchanging 
the components of $\alpha$ and $\beta$ correspond to exchanging rows or columns of $S$, 
which does not change its singular values. This implies that the entries of 
$\ket{\alpha}\bra{\beta}$ decrease in each row and column. Different values for 
$\alpha_i$ and $\beta_i$ thus lead to different arrangements. Consequently, we have 
to optimize $S$ under the constraints that the entries of $S$ must be non-increasing 
in each row from left to right and in each column from top to bottom.

%
%From the theory of Young tableaux \cite{Georgi} it follows that under these 
%constraints for a decomposition into $d=d_1\times d_2$ there 
%are at most
%Furthermore, if one is only interested in decomposability, it suffices 
%to check whether there exists an arrangement such that $S$ has rank one. 
%This adds further restrictions since the rows and columns must be linearly dependent. 
%The number of possible arrangements reduces to at most
%(see Appendix C \cite{appremark})

\begin{figure}[t]
\centering
\includegraphics[scale=.3]{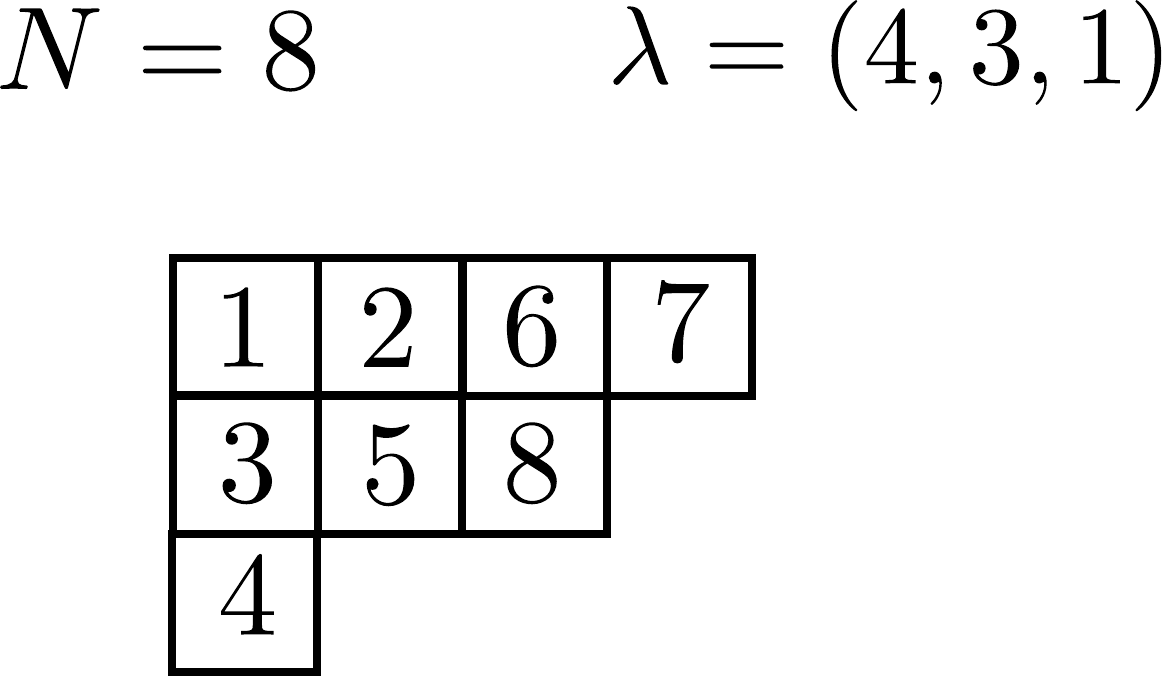}
\caption{This figure shows an example of a standard Young tableau for $N=8$ and a partitioning $\lambda=(4,3,1)$. The numbers $1,\dots,8$ are arranged in such a way that their values increase in each row and each column.}
\label{Fig:Young}
\end{figure}

To see the connection to Young tableaux, let us first recall the definition of a 
Young diagram. Given some number $N\in\mathbb{N}$ we call 
$\lambda=(\lambda_1,\lambda_2\cdots,\lambda_n)$ a partitioning of the 
number $N$, that is $\sum_k\lambda_k=N$, $\lambda_1\geq\lambda_2\geq\cdots\geq\lambda_n$, 
and $\lambda_i\in \mathbb{N}$. Then a Young diagram is an 
arrangement of left-justified rows, where the number of boxes 
in the $k$-th row is given by $\lambda_k$ (see Fig.~\ref{Fig:Young}). 

A Young tableau of shape $\lambda$ is a filling of the numbers 
$1, 2, \cdots, n$ into the boxes of the Young diagram such 
that every number appears exactly once. A Young tableau 
is called standard if the numbers are increasing in each 
row and each column. From here it is straightforward to see 
that this problem is equivalent to the problem of finding the 
number of possible arrangements of the Schmidt coefficients 
in the matrix $S$ under the constraints that we discussed above.
The number of possible arrangements that could lead to different 
maximal singular values is simply given by the number of standard
Young tableaux consisting of $d_1\times d_2$ boxes, arranged in a 
single block. This number is given by the so-called hook-length 
formula \cite{Georgi}
\begin{equation}
\mathcal{N}=\frac{n!}{\prod_{(i,j)}h_{i,j}},
\end{equation}
where $h_{i,j}$ is called a hook-length of the box $(i,j)$. For a given 
box $(i',j')$, its hook consists of all boxes with either $(i=i',j>j')$ 
or $(i>i',j=j')$ and the box itself. The length of the hook is then given
by the number of boxes in the hook. For a Young tableau of $d_1\times d_2$ 
boxes this simplifies to
\begin{equation}
\mathcal{N}=\frac{(d_1\times d_2)!}{\prod_{i=1}^{d_1}\prod_{j=1}^{d_2}(i+j-1)}.
\end{equation}

In case one is only interested whether or not a state is decomposable,
the number of different matrices that lead to a maximal singular value 
of one can be further reduced. This is due to the additional constraint 
that if the matrix $S$ has rank one all the rows as well as the columns 
must me mutually linearly dependent. Then, it is easy to see that the 
following algorithm can solve the problem. We start again by filling 
the Schmidt coefficients in an array such that their values are
non-increasing in each row and each column. We can fix the upper 
left entry to be the largest element. Whenever we get in a situation 
in which we fix the constant between two rows or columns we check
whether there are some remaining Schmidt coefficients which lead 
to linearly dependent rows or columns. If this is the case, 
we fill the array with the appropriate number and continue. 
If these numbers do not exist, we abort and have to start all 
over again with a different arrangement. It is obvious that 
if there exists an 
arrangement which leads to a matrix with rank one, then the 
algorithm will find it. Using the formalism of Young tableaux 
we can again calculate the maximum number of different matrices
that we need to check. First note that when we apply the algorithm 
we always fix the values of the entries in the first row and 
the first column. The only thing that changes is the order in 
which we fill the entries. The number of all possible ways 
to do this is again given by a number of standard Young 
tableaux consisting of a single row and a single column.
By applying the hook length formula we obtain
\begin{eqnarray}
\mathcal{N'}&=&\frac{(d_1+d_2-1)!}{(d_1+d_2-1)\times (d_1-1)!\times (d_2-1)!} \notag\\
&=&\frac{(d_1+d_2-2)!}{(d_1-1)!\times (d_2-1)!}.
\end{eqnarray}

%%%%%%%%%%%%%%%%%%%%%%%%%%%%%%%%%%%%%%%%%%%%%%%%%%%%%%%%%

\subsection{D: Examples}
In this section, we provide some notes on Example 1 
(fully decomposable state) as well as  a detailed proof 
for Example 2 (MME state) for the multipartite exemplary 
states given in the main text. Furthermore we present another 
interesting fully decomposable state of six qubits.

\subsubsection{Example 1. GHZ States}

\paragraph{LU-equivalence of GHZ- and star-type graph states.}
Here we show the equivalence of star-type graph states and GHZ 
states in arbitrary dimension and system size under local unitary 
(LU) operations. Decomposability is a property of a state which 
does not change under LU-operations on the original state, therefore 
it is sufficient to show that $\ket{G_{star}} \stackrel{LU}{=} \ket{GHZ^{(D)}}$ 
for any dimension $D$ and any number of qudits $N$.

Star-type graphs are graphs where one central vertex is connected to any other 
vertex by an edge, and no other edges are present. For the corresponding quantum
state we have according to Eq.~(\ref{eq:graphstatedef}) in the main text
$\ket{G_{star}}= \prod_{q=2}^N Z_{1q} \ket{+}^{\otimes N}$.
This can be simplified to: 
\begin{equation}
\label{stargraph}
\begin{split}
\ket{G_{star}} &=
\sum_{p=0}^{N-1}  
\ket{p}_1 
\bigotimes_{q=2}^N
\ket{+_p}_q 
\\
&\propto \ket{0}_1\ket{+_0}_2 \dots \ket{+_0}_N + \ket{1}_1\ket{+_1}_2
\dots \ket{+_1}_N
\\
& \quad +\dots +\ket{D-1}_1\ket{+_{D-1}}_2 \dots \ket{+_{D-1}}_N.
\end{split}
\end{equation}
Here we use the ($D$-dimensional) single qudit states 
$\ket{+_i}={1\over {\sqrt{D}}} \sum_{k=0}^{D-1} \omega^{ki}\ket{k}$ 
with $\omega = e^{2 \pi i/D}$, note that $\ket{+_0} = \ket{+^D}$ in
our previous notation. Since $\braket{+_i}{+_j}=\delta_{ij}$ the set $\{\ket{+_i}\}$  forms a basis of $\mathbb{C}^D$. Eq.~\eqref{stargraph} 
is, up to local rotations on all subsystems except the first, equal to 
$\ket{GHZ^{(D)}}$.

\paragraph{Full decomposability of a $6\times 6\times 6$ system}
To clarify the proof of Eq.~\eqref{GHZ} in the main text, we exemplary 
do the complete calculation for a system of three parties each of which 
has dimension six, such that the prime decomposition $D=2\times3$ equals
access to a qubit and a qutrit.
The state, up to normalization, reads $\ket{GHZ^{(6)}}=\sum_{\ell=0}^5 \ket{\ell \ell \ell}$.
The encoding and resorting of the order, which groups the subsystems of the 
qubits and qutrits respectively, then gives the six-partite state:
\begin{equation}
 \begin{split}
  \ket{GHZ^{(6)}} &\stackrel{enc.} {=} \ket{000000}+\ket{010101}+\ket{020202}\\
  &~~~~~~~+\ket{101010}+\ket{111111}+\ket{121212}\\
  & \stackrel {res.} {=} \ket{000000}+\ket{000111}+\ket{000222}\\
  &~~~~~~~+\ket{111000}+\ket{111111}+\ket{111222}\\
  &\stackrel{\phantom{enc.}}= (\ket{000}+\ket{111})\otimes(\ket{000}+\ket{111}+\ket{222})
 \end{split}
\end{equation}
which shows decomposability into $\ket{GHZ^{(2)}}$ and $\ket{GHZ^{(3)}}$
The generalization to an arbitrary number of systems $N$ 
and arbitrary dimension $D$ follows straightforward.

%%%%%%%%%%%%%%%%%%%%%%%%%%%%%%%%%%%%%%%%%%%%%%%%%%%%%%%%%%%%%%%

%%%%%%%%%%%%%%%%%%%%%%%%%%%%%%%%%%%%%%%%%%%%%%%%%%%%%%%%%%%%%%%

 \subsubsection{Example 2: Graph states}
\begin{figure} 
     \centering
     \includegraphics[width=0.4\textwidth]{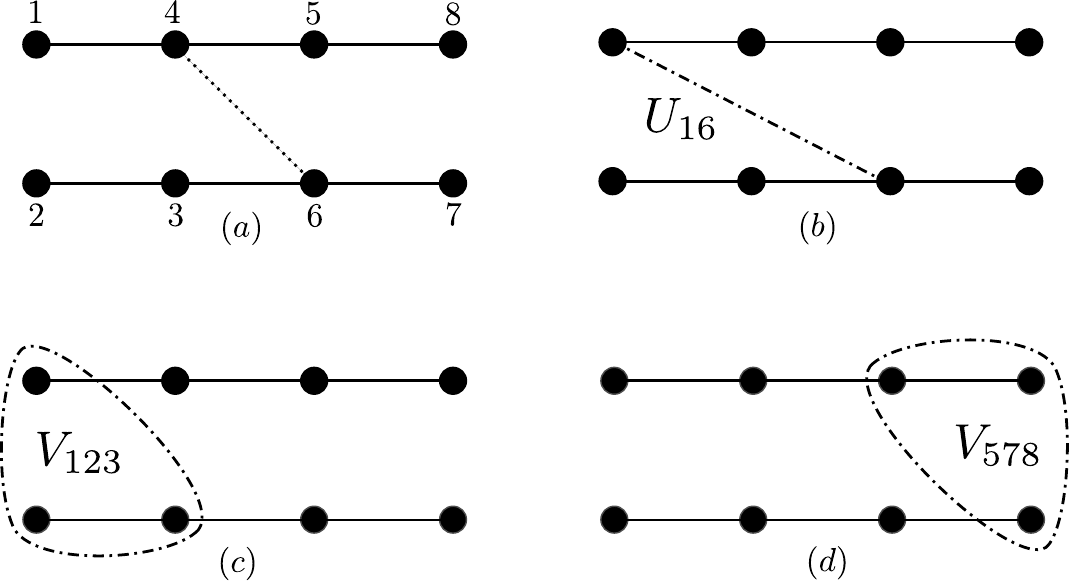}
     \caption{
     Example of a state that is MME but not genuine MME. The four-ququart 
     chain-type graph is encoded into LU-equivalent eight-qubit states.
     (a): The equivalence to this state has already been shown in the main 
     text, see Fig.~\ref{fig:mme}. (b): The state is also equivalent to this
     configuration, see Eq.~(\ref{chain0B3-app}). (c) and (d): These equivalences
     follow from Eq.~(\ref{chainQB2-app}). In summary, the state is is 
     decomposable with respect to all possible bipartitions.}
     \label{fig:MME_HG}
 \end{figure}

Here we present the calculation for the four-ququart graph states, 
see also Fig.~\ref{fig:MME_HG} and Fig.~\ref{fig:mme} in the main text. 
To start, the chain graph state of $N=4$ ququarts is given by
 \begin{equation} 
 \label{QQchain-app}
 \begin{split}
  \ket{G^{(4)}}&= \tilde{Z}_{AB}\tilde{Z}_{BC}\tilde{Z}_{CD}\ket{+^{4}}^{\otimes 4}\\
  &=\sum_{ABCD=0}^3
  \omega_{(4)}^{AB} \omega_{(4)}^{BC}
  \omega_{(4)}^{CD}\ket{ABCD}.
  \end{split}
 \end{equation}
 Here $\tilde{Z}_{ij}=\text{diag}(1,i,-1,-i)$ is the ququart controlled Z-gate
 and $\omega_{(4)} = e^{2 \pi i / 4} =i.$
 We use the computational basis $\ket{ABCD}$ to simplify the encoding process. The ququarts corresponding to $(A,B,C,D)$ are decomposed into two qubits each 
 with the labels
 $(A_1,A_2,B_1,B_2,C_1,C_2,D_1,D_2)=(1,2,4,3,5,6,8,7)$, see 
 Fig.~\ref{fig:MME_HG}(a).

 To represent the ququart state, 
 we make the replacements: 
 $A \rightarrow 2A_1+A_2$, $B \rightarrow 2B_2+B_1$, $C \rightarrow 2C_1+C_2$ 
 and $D \rightarrow 2D_2+D_1$,
 as this  reproduces for an additional replacement of the $\sum_{A,B,C,D=0}^3 \rightarrow
 \sum_{A_1...D_2=0}^1$ the same exponents as in Eq.~(\ref{QQchain-app}).
 Then we have:
 \begin{equation} \label{QBchain1-app}
 \begin{split}
  \ket{G^{(4)}}&\stackrel {enc.}= \ket{G^{(2)}}\\
  &\stackrel {\phantom{enc.}}=\sum_{A_1...D_2=0}^1 \omega_{(4)}^{(2A_1+A_2)(2B_2+B_1)}
  \omega_{(4)}^{(2B_2+B_1)(2C_1+C_2)}\\
  &~~~~~~~~~~~\omega_{(4)}^{(2C_1+C_2)(2D_2+D_1)}\ket{A_1A_2B_1B_2C_1C_2D_1D_2}.
  \end{split}
 \end{equation}
We furthermore use $\omega_{(4)}=e^{{i\pi} \over 2}=\omega_{(2)}^{1\over 2}$
and $\omega_{(2)}^{2c}=1,~c \in \mathbb{N}$ and can simplify 
Eq.~\eqref{QBchain1-app}
\begin{equation}\label{QBwe-app}
\begin{split}
 \ket{G^{(2)}}&=\sum_{A_1...D_2=0}^2 \omega_{(2)}^{A_1B_1}
 \omega_{(2)}^{A_2B_2}\omega_{(2)}^{B_2C_2}\omega_{(2)}^{B_1C_1}
 \omega_{(2)}^{C_1D_1}\omega_{(2)}^{C_2D_2}\\
 &~~~~~~~~~~~~~\omega_{(2)}^{{A_2B_1} \over 2} \omega_{(2)}^{{B_1C_2} \over 2}
 \omega_{(2)}^{{C_2D_1} \over 2}\ket{A_1A_2B_1B_2C_1C_2D_1D_2}\\
 &=Z_{A_1B_1}Z_{A_2B_2}Z_{B_2C_2}Z_{B_1C_1}Z_{C_1D_1}Z_{C_2D_2}\\
 &~~~~~~Z_{A_2B_1}^{1\over 2}Z_{B_1C_2}^{1\over 2}Z_{C_2D_1}^{1\over 2}\ket{+^{2}}^{\otimes 8}.
 \end{split}
 \end{equation}
 Here, $Z_{ij}={\rm diag}(1,-1)$ is the qubit-controlled Z-gate, this 
 state is shown in left side of Fig.~\ref{fig:mme} in the main text.
 We then apply  $V_{A_1A_2}$, $V_{B_1B_2}^{3 \over 2}$
and  $V_{D_1D_2}^{3 \over 2}$ to $\ket{G^{(2)}}$.
Those are for the further analysis in this example defined as 
 \begin{equation}
 \label{UNITV-app}
     V_{X_1X_2}=(\ket{+}\bra{+})_{X_1} \otimes \openone_{X_2}+
 (\ket{-}\bra{-})_{X_1} \otimes Z_{X_2},
 \end{equation}
where for $X=A,B,C,D$ all $V_{X_1X_2}$ are included in the 
set of vertical unitaries $\{U_{\rm Vert}\}$.
By straightforward calculation, one verifies:
 \begin{equation}
 \label{chainQB2-app}
 \begin{split}
 ( V_{A_1A_2} V_{B_1B_2}^{3 \over2}V_{D_1D_2}^{3\over 2}) \ket{G^{(2)}}&=
 V_{A_1A_2B_2}\ket{G^{(2)}_{\DD}},\\
  ( V_{D_1D_2} V_{C_1C_2}^{3 \over2}V_{A_1A_2}^{3\over 2}) \ket{G^{(2)}}&=
 V_{C_1D_1D_2}\ket{G^{(2)}_{\DD}}.
 \end{split}
 \end{equation}
This means that for the question of decomposability the weighted diagonal 
edges have the same effect on the decomposable state $\ket{G^{(2)}_{\DD}}$ 
as one hyper-edge connecting  three qubits either one or the other end of 
the chain, see Fig.~\ref{fig:MME_HG}(c) and Fig.~\ref{fig:MME_HG}(d).
The mentioned hyper-edge is formally a three-qubit unitary of the form
\begin{equation}
V_{X_1X_2Y_1}=(\ket{+}\bra{+})_{X_1} \otimes \openone_{X_2 Y_1}+ (\ket{-}\bra{-})_{X_1} \otimes V_{X_2 Y_1}
\label{chainQB4-app}
\end{equation} 
with $V_{X_2 Y_1}$ as defined in Eq.~\eqref{UNITV-app} 
and $\ket{G^{(2)}_{\DD}}$ is a decomposable
state, defined in Eq.~(\ref{bidecgraph}) in the main text.

 Furthermore, one can directly check that we can replace the three weighted 
 Z-gates ($Z^{1 \over 2}_{ij}$) in Eq.~\eqref{QBwe-app} by one 
 weighted edge acting on qubits $A_1$ and $C_2$
 \begin{equation}
 \label{chain0B3-app}
  \ket{G^{(2)}}= U_{A_1A_2}^{3\over 2} U_{B_1B_2}^{3\over 2}U_{A_1C_2} \ket{G^{(2)}_{\DD}}
 \end{equation}
with $U_{A_1C_2}=(\ket{+}\bra{+})_{A_1}\otimes \openone_{C_2}+
 (\ket{-}\bra{-})_{A_1} \otimes Z_{C_2}$ and two vertical unitaries $U_{A_1A_2}^{3\over 2}$ and $ U_{B_1B_2}^{3\over 2}$ 
 [see Fig.~\ref{fig:MME_HG}(b)].

 From Eq.~\eqref{chainQB2-app} and Eq.~\eqref{chain0B3-app} we see that 
 whereas this state is not decomposable, there exists for every bipartition 
 a representation of this state, for which the $S$-matrix has rank 1. 
 In Fig.~\ref{fig:MME_HG}, the different equivalent representations of the state are shown graphically. Each option presents decomposability with respect to another bipartite split, such that all possible ones are covered.
 However, to exclude genuine MME, let us once again stress that the existence of one split exhibiting decomposability is enough.

%%%%%%%%%%%%%%%%%%%%%%%%%%%%%%%%%%%%%%%%%%%%%%%%%%%%%%%%%%%%%%%%%%%%%%%5

%%%%%%%%%%%%%%%%%%%%%%%%%%%%%%%%%%%%%%%%%%%%%%%%%%%%%%%%%%%%%%%%%%%%%%%5
\begin{figure}[t]
\begin{center}
 \includegraphics[width=0.9\columnwidth]{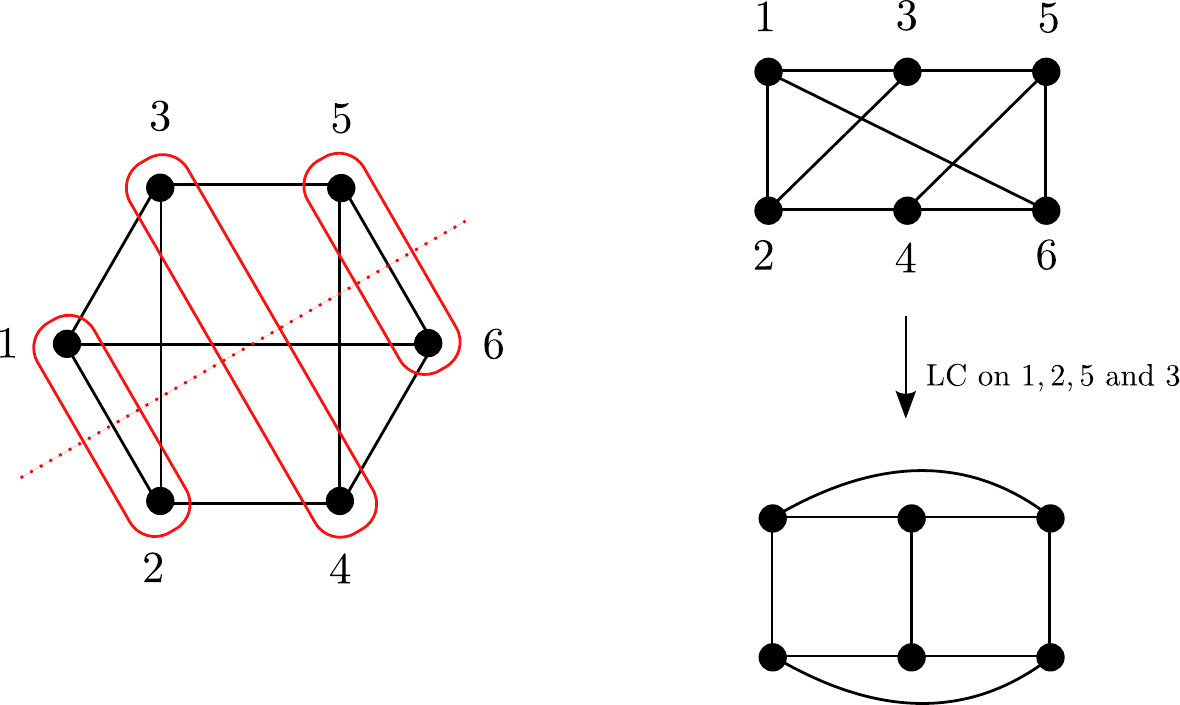}
 \end{center}
 \caption{The maximally entangled state on six qubits represents a 
 decomposable three ququart state.}
 \label{fig:6qubitmax}
\end{figure}
%%%%%%%%%%%%%%%%%%%%%%%%%%%%%%%%%%%%%%%%%%%%%%%%%%%%%%%%%%%%%%%%%%%%%%%%%

\subsubsection{The maximally entangled state of six qubits}
We have already seen that the highly entangled GHZ states are not necessarily multilevel entangled. Therefore one might ask the following question: Are there other highly entangled states which are not multilevel entangled? One example is the three-ququart state that corresponds to the absolutely maximally entangled state of six qubits (see Fig.~\ref{fig:6qubitmax}). The six-qubit state is 
given by
\begin{equation} 
\label{FD1}
 \ket{G^{(2)}}=Z_{12}Z_{34}Z_{56}Z_{23}Z_{36}Z_{45}Z_{24}Z_{35}Z_{16}\ket{+}^{\otimes 6}
\end{equation}
and corresponds to a graph state.
Nevertheless, this state is fully decomposable. 
To prove this, we first mention that via local complementation \cite{weighted_graphs_hein} (LC), we can obtain:
\begin{equation}\label{FD2}
 \ket{G^{(2)}} \xrightarrow[\text{on 1,2,5,3}]{\text{LC}} Z_{12}Z_{56}Z_{14}Z_{23}Z_{36}Z_{45}Z_{15}Z_{26} \ket{+}^{\otimes 6}
\end{equation}
Comparing Eq.~\eqref{FD1} and Eq. \eqref{FD2}, the difference between 
those is depicted in Fig.~\ref{fig:6qubitmax} on the right side. Whereas 
the first contains diagonal connections (which contradicts a direct 
decomposition), the second form shows that these can be replaced 
by vertical and horizontal ones. Therefore we can reach the original 
state by starting from a decomposable state.

%%%%%%%%%%%%%%%%%%%%%%%%%%%%%%%%%%%%%%%%%%%%%%%%%%%%%%%%%%%%%%%%%%%%%%%

%%%%%%%%%%%%%%%%%%%%%%%%%%%%%%%%%%%%%%%%%%%%%%%%%%%%%%%%%%%%%%%%%%%%%%%%%

%\subsubsection{A bidecomposdable, but not fully decomposable state}

%Finally, let us give an example of a pure three-ququart state that 
%is bidecomposable for any bipartition, but where one can prove 
%analytically that the state is not fully decomposable. Note that 
%this phenomenon does not occur in usual entanglement theory: A 
%pure state that is separable for any bipartition is also fully 
%separable.

%Consider the state
%\begin{equation}
%\ket{\psi} = \frac{1}{\mathcal{N}}(
%\ket{0} \ket{0} \ket{\phi_0}+
%\ket{1} \ket{1} \ket{\phi_1}+
%\ket{2} \ket{2} \ket{\phi_2})
%\end{equation}
%where $\mathcal{N}$ denotes a normalization. The $\ket{\phi_i}$ 
%are assumed to be nearly the same, formalized as 
%$|\braket{\phi_i}{\phi_j}|^2 \geq 1 - \varepsilon$, 
%moreover, let us assume that the three $\ket{\phi_i}$ 
%span only a two-dimensional space.

%Then, the resulting state $\ket{\psi}$ has the desired properties: 
%Since the $\ket{\phi_i}$  are linearly dependent, the Schmidt rank 
%for the $AB|C$ bipartition is two and the state is bidecomposable 
%with respect to that partition. However, the reduced state \rho_AB is nearly
%the maximally entangled qutrit state. it is not a convex combination
%of decomposable states (see the witness in our paper). From
%this it follows, that there cannot be UA and UB that map rho_AB
%to  rho_A1B1 \otimes rho_A2B2. which shows that the state is not
%fully decomposable.
%------------------------- 
\subsection{E: Algorithm for testing full decomposability}
\begin{figure}[t]
 \centering
 \includegraphics[scale=.7]{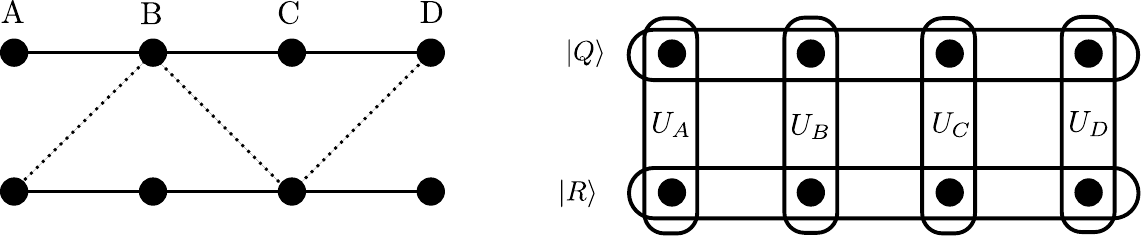}
 \caption{We ask whether the state on the left can be constructed by first preparing states $\ket{Q}$ and $\ket{R}$ and then applying local unitary operations $U_A,\dots, U_D$. 
 Since this is not possible the state is not decomposable, it is MME.}
 \label{fig:6qubitopt}
\end{figure}

In this section we explain the algorithm that we used to test 
whether or not the four ququart chain-graph state $\ket{\psi}$ 
in Eq.~\ref{QQchain} in the main text. The aim is to test whether 
or not the state $\ket{\psi}$ can be written 
as 
$\ket{\psi}\stackrel{\text{?}}{=}U_A\otimes U_B\otimes U_C\otimes U_D \ket{Q}\otimes\ket{R}$, 
see also Fig.~\ref{fig:6qubitopt}. Thus, we want to compute
\begin{equation}
\max_{\stackrel{U_A\cdots U_D}{\ket{Q}\ket{R}}}\abs{\bra{Q}\bra{R}U_A\otimes U_B\otimes U_C\otimes U_D\ket{\psi}}.
\end{equation}
The idea is to choose initial states $\ket{Q}$ and $\ket{R}$, as well 
as unitaries $U_A,\dots,U_D$ at random and then optimize the states 
and unitaries iteratively, until a fix-point is reached. The point is that
any of the iteration steps can be performed analytically. 
In order to optimize the state 
$\ket{Q}$, we fix the unitaries $U_A,\dots,U_D$ and the state $\ket{R}$. 
We obtain the optimal choice of $\ket{Q}$ by computing 
$\max_{Q}\abs{\bra{Q}(\bra{R}U_A\otimes U_B\otimes U_C\otimes U_D\ket{\psi})}=\max_{Q}|\braket{Q}{\tilde{\psi}}|$. We have 
that $\ket{Q}\propto\ket{\tilde{\psi}}$ is optimal up to normalization. 
The similar argument holds for $\ket{R}$. For optimizing the local 
unitaries we fix any unitary but the one we want to optimize, 
say $U_A$. Then, we have
\begin{eqnarray}
&& \max_{U_A}\abs{\bra{Q}\bra{R}U_A\otimes U_B \otimes
U_C\otimes U_D\ket{\psi}}
\notag \\
&=&\max_{U_A}\abs{\bra{Q}\bra{R}U_A\tilde{\ket{\psi}}}\notag \\
&=&\max_{U_A}\abs{\tr( U_A\tilde{\ket{\psi}}\bra{Q}\bra{R})}\notag \\
&=&\max_{U_A}\abs{\tr_A(U_A\varrho_A)}=\sum_i s_i(\varrho_A)
\end{eqnarray}
where $\varrho_A=\tr_{BCD}(\tilde{\ket{\psi}}\bra{Q}\bra{R})$. We write
$\varrho_A$ in the singular value decomposition and we get
$\varrho_A=UDV^{\dagger}$. Then we choose $U_A=VU^{\dagger}$ and hence
\begin{equation}
\abs{\tr_A(U_A\varrho_A)}=\abs{\tr(D)}=\sum_i s_i(\varrho_A).
\end{equation}

%%%%%%%%%%%%%%%%%%%%%%%%%%%%%%%%%%%%%%%%%%%%%%%%%%%%%%%%%
%%%%%%%%%%%%%%%%%%%%%%%%%%%%%%%%%%%%%%%%%%%%%%%%%%%%%%%%%

\end{document}